\documentclass[12pt]{iopart}
\def\beqra{\begin{eqnarray}}
\def\eeqra{\end{eqnarray}}
\def\beq{\begin{equation}}
\def\eeq{\end{equation}}

\begin{document}

\title{Cosmological implications of a supersymmetric extension of the Brans-Dicke theory}

\author{Riccardo Catena}

\address{Deutsches Elektronen-Syncrotron DESY, 22603 Hamburg, Germany}

%\author{Riccardo Catena} 
%\affiliation{Deutsches Elektronen-Syncrotron (DESY), 
%\\ 22603 Hamburg, Germany \\ {\tt (catena@mail.desy.de)}}

\begin{abstract}
In the Brans-Dicke theory the Planck mass 
is replaced by a dynamical scalar field. 
We consider here the supersymmetric analogous of this mechanism
replacing in the supergravity Lagrangian the Planck mass
with a chiral superfield.
This analysis is motivated by  
the research of possible connections between supersymmetric Dark Matter scenarios and
Dark Energy models based on Brans-Dicke-like theories. 
We find that, contrary to the original Brans-Dicke theory, in its supersymmetric analogous
the gravitational sector does not couple to the matter sector in a universal metric way. 
As a result, violations of the weak equivalence principle
could be present in such a scenario.
\end{abstract}

DESY 07 -- 143
%\maketitle

%%%%%%%%%%%%%%%%%%%%%%%%%%%%%%%%%%%%%%%%%%%%%%%%%%%%
\section{Introduction}
%%%%%%%%%%%%%%%%%%%%%%%%%%%%%%%%%%%%%%%%%%%%%%%%%%%%

In the Brans-Dicke approach to the gravitational interaction, the Planck mass 
is replaced in the Lagrangian by a dynamical scalar field $\varphi$~\cite{BD}. 
Even though, as a theory of gravity,
the original Brans-Dicke theory is at present very constrained by solar system measurements~\cite{Cassini}, 
its modern versions, namely Scalar-Tensor (ST) theories~\cite{ST} -- where a non trivial potential 
is associated to $\varphi\,$ --  
can still pass all cosmological and astrophysical bounds leading at the same time to
interesting and testable predictions~\cite{predictions}.

Like in General Relativity, in ST theories matter couples to gravity in 
a universal metric way
{\it i.e.} all matter 
fields feel the gravitational interaction only through {\it one and the same} metric. 
Because of this property, ST theories are by construction protected by 
any violation of the weak equivalence principle and even  
ultra-light degrees of freedom mediating long range forces are not in this framework
phenomenologically dangerous~\cite{Dam}. Such a feature makes ST theories interesting
arenas for dynamical Dark Energy model building\cite{STDE,DE}. 

However, a realistic cosmological scenario can be only achieved when
Dark Matter is consistently included to the picture~\cite{WMAP}. Among different theories providing
suitable Dark Matter candidates, Supersymmetry is certainly one of the most remarkable~\cite{DM}.

The interesting possibility to relate a ST interpretation
of Dark Energy to a supersymmetric description of Dark Matter leads
to study supersymmetric extensions of ST theories.
This is the topic of the present work.

We consider here the supersymmetric analogous of the BD idea:
we replace in the supergravity Lagrangian the Planck mass
with a chiral superfield, the ``Planck superfield''.
Such a replacement defines the ``natural'' supersymmetric extension of the 
BD theory. Let us refer to it as 
the Minimal Supersymmetric Brans-Dicke theory (MSBD) to distinguish it
from other possible approaches.
We find that, contrary to the original BD theory, in the MSBD
the gravitational sector
does not couple to the matter sector in a universal metric way. 
As a result, possible violations of the weak equivalence principle
could make the minimal supersymmetric extension of the 
BD idea phenomenologically inconsistent.
 
The plan of this work is as follows. 
In the next section we review 
the BD model and the concept of universal metric coupling. 
The third section is devoted to the MSBD theory;
we will specially underline the differences between its phenomenology and the one of the original BD theory.  
The results are finally discussed in the Conclusions.
For notation we refer in the following to~\cite{Wess,Riccardo}.
%\section{Notation}
%\label{not}
%We will use in the following the same notation and conventions of \cite{Wess}. 
%We list here for clarity some of them.
%
%The superspace is described in terms of the coordinates $(y^m, \theta_{\alpha})$.
%Greek indexes label two components Weyl spinors while latin indexes the components of four-vectors.
%Indexes transforming under local coordinates transformations in superspace are called Einstein indexes
%and are taken from the end of the alphabet, for example $(m, \,n, \, \dots)$. 
%Instead, indexes transforming under local Lorentz transformations are called Lorentz indexes and
%are taken from the beginning of the alphabet, for example $(a, \,b, \, \dots)$. 
%The power series expansion in $\theta_{\alpha}$ of a chiral superfield $\Phi$ is given by
%%
%\beq
%\Phi(y^m,\theta_{\alpha}) = A(y^m) + \sqrt{2} \,\theta^{\alpha} \chi_{\alpha}(y^{m}) + \theta^{\alpha} \theta_{\alpha} F(y^{m})
%\label{ps}
%\eeq  
%%
%where $A(y^{m})$ and $F(y^{m})$ are complex scalars and $\chi_{\alpha}(y^{m})$ a Weyl spinor.
%We will couple matter superfields to the minimal supergravity multiplet.
%This contains the vielbein $e^a_{m}$, the gravitino $\psi^a_{\alpha}$ and two auxiliary fields:
%a vector $b^a$ and a scalar $M$. Finally, covariant derivatives with respect to
%supergravity transformations are denoted by $\mathcal{D_{\alpha}}$, $\bar{\mathcal{D}}^{\dot{\alpha}}$ and  $\mathcal{D}_{m}$.
%
%
%%%%%%%%%%%%%%%%%%%%%%%%%%%%%%%%%%%%%%%%%%%%%%%%%%%%%%%%%%%%%%%%%%%%%%%%%%%%%%%%%%%%%%%%%%%%%%%%%
\section{The Brans-Dicke theory and the universal metric coupling}
\label{BDumc}
%%%%%%%%%%%%%%%%%%%%%%%%%%%%%%%%%%%%%%%%%%%%%%%%%%%%%%%%%%%%%%%%%%%%%%%%%%%%%%%%%%%%%%%%%%%%%%%%%

In General Relativity the coupling between gravity and matter is described by the following
Lagrangian
\beq
\mathcal{L}_{EH} = -\frac{1}{2} \,e M_{Pl}^2 \mathcal{R}  + \mathcal{L}_{M}[e^{a}_{m}, \Psi] \,,
\label{eh}
\eeq
where $e\equiv det(e^{a}_{m})$, $\mathcal{R}$ is the Ricci scalar and $\Psi$ symbolically
represents all matter fields involved in the theory.
In the BD approach to the gravitational interaction the Planck mass appearing in eq.~(\ref{eh}) becomes dynamical 
by means of the substitution
\beq
M_{Pl}^2 \Longrightarrow \varphi^{2}(y^m) \,,
\label{sp}
\eeq
where $\varphi(y^m)$ is a real scalar field.
As a consequence eq.~(\ref{eh}) is replaced by
\beqra
\mathcal{L}_{BD} &=& \mathcal{L}_{\varphi}[e^{a}_{m}, \varphi] + 
\mathcal{L}_{M}[e^{a}_{m}, \Psi] \nonumber\\
&=& - \frac{1}{2} \,e \left( \varphi^2  \, \mathcal{R}  + \omega\, \partial_{m} \varphi \partial^{m} \varphi \right) 
+ \mathcal{L}_{M}[e^{a}_{m}, \Psi] \,,
\label{bd}
\eeqra
where the factor $\omega$ that multiplies the kinetic term of $\varphi$ has to be tuned
to fit the post-newtonian bounds \cite{Cassini}. 
%\footnote{At present, the stronger bound, $\omega > 10^4$, 
%comes from time-delay measurements of the Cassini spacecraft.}
Eq.~(\ref{bd}) gives the so called ``Jordan frame'' formulation of the theory.
In this frame the BD scalar does not appear in the matter Lagrangian and particle physics is just the standard one.
The theory can be formulated in other frames related to the Jordan one by a Weyl rescaling
of the vielbein such as $e^{a}_{m} \to  e^{a}_{m}\, e^{l(\varphi)} $, where $l(\varphi)$ is some 
$\varphi$-dependent function. 
In these alternative formulations the matter Lagrangian acquires
an explicit functional dependence from $\varphi$, {\it i.e.} 
$\mathcal{L}_{M} = \mathcal{L}_{M}[e^{a}_{m} \, e^{l(\varphi)} ,\Psi]$.  
However, the inverse Weyl rescaling $e^{a}_{m} \to  e^{a}_{m}\, e^{-l(\varphi)} $
always brings back the theory to its original version in which
particle physics is just the standard one.

Eq.~(\ref{bd}) shows that in the BD theory all matter fields feel the gravitational interaction
through the same vielbein, the Jordan frame vielbein. For this reason such a matter-gravity
coupling is also called {\it universal and metric}.  
This is a non trivial property and has very important phenomenological implications.
It can be shown, for instance, that in a theory where matter couples to gravity in a universal metric
way the weak equivalence principle is satisfied by construction \cite{Dam}.
\section{The Minimal Supersymmetric Brans-Dicke theory}
%\section{A supersymmetric Brans-Dicke theory}
\label{susyBD}
%%%%%%%%%%%%%%%%%%%%%%%%%%%%%%%%%%%%%%%%%%%%%%%%%%%%%%%%%%%%%%%%%%%%%%%%%%%%%%%%%%%%%%%%%%%%%%%%%

Eq.~(\ref{sp}) gives a prescription to construct the BD Lagrangian starting from the Einstein-Hilbert one.
In this section we apply an analogous prescription to the supergravity Lagrangian
\beq
\mathcal{L}_{sg} = -3\,M_{Pl}^{2} \int d^2\theta \, 2 \mathcal{E} R + \mathcal{L}_{M}[H, \Psi] + \textrm{h.c.} \,, 
\label{sg}
\eeq
where $H$ is the supergravity multiplet,
$\mathcal{E}$ is the chiral density and
$R$ represents the curvature superfield, defined as the covariant derivative of the spin connection.\\
Let us start introducing a chiral superfield $\Phi$ 
with components given by the power series expansion $\Phi(y^m,\theta_{\alpha}) = A(y^m) + \sqrt{2} \,\theta^{\alpha} \chi_{\alpha}(y^{m}) + \theta^{\alpha} \theta_{\alpha} F(y^{m})$,
%
%\beq
%\Phi(y^m,\theta_{\alpha}) = A(y^m) + \sqrt{2} \,\theta^{\alpha} \chi_{\alpha}(y^{m}) + \theta^{\alpha} \theta_{\alpha} F(y^{m})
%\label{ps}
%\eeq  
%
where $A(y^{m})$ and $F(y^{m})$ are complex scalars and $\chi_{\alpha}(y^{m})$ a Weyl spinor.
We will call $\Phi$ the Planck superfield. This dynamical object allows
the natural supersymmetric extension of the substitution~(\ref{sp})
\beq
M_{Pl}^2 \Longrightarrow \Phi^{2}(y^m,\theta_{\alpha}) \,.
\label{sr}
\eeq

Applying the substitution~(\ref{sr}) to eq.~({\ref{sg}}) one finds 
\beqra
\mathcal{L}_{MSBD} &=& \mathcal{L}_{\Phi}[H, \Phi] + \mathcal{L}_{M}[H, \Psi] \nonumber\\ 
&=& -3 \int d^2\theta \, \Phi^2 \, 2 \mathcal{E} R - \nonumber\\
&-&\frac{1}{8} \int d^2\theta \, 2 \mathcal{E}  \left(\bar{\mathcal{D}}_{\dot{\alpha}} \bar{\mathcal{D}}^{\dot{\alpha}} 
-8\,R \right) \Phi^{\dagger} \Phi  + \nonumber\\
&+& \mathcal{L}_{M}[H,\Psi]  \,+\, \textrm{h.c.} \,, 
\label{SBD}
\eeqra
where in the third line, in  analogy with eq.~(\ref{bd}), we introduced a kinetic term for $\Phi$.
To be as general as possible we do not assume any particular form for $\mathcal{L}_{M}$.

Eq.~(\ref{SBD}) defines the Minimal Supersymmetric Brans Dicke theory (MSBD).
Its invariance under supergravity transformations follows from the properties of chiral
densities. By definitions, chiral densities transform like total derivatives
in the space $(y^{m}, \theta_{\alpha})$ and the product of a chiral density and a chiral
superfield is again a chiral density \cite{Wess}. Moreover, the superfields $\left(\bar{\mathcal{D}} \bar{\mathcal{D}} 
-8\,R \right) \Phi^{\dagger} \Phi$ and $\Phi^{2}$ are chiral if $\Phi$ is chiral.
This proves the invariance of the Lagrangian~(\ref{SBD}) under supergravity transformations.

Let us focus now on its phenomenology. As it was shown in~\cite{Riccardo}, 
the component fields expansion of eq.~(\ref{SBD}) gives rise to a Lagrangian with the following structure 
\beq
\mathcal{L}_{MSBD} = \mathcal{L}_{\Phi}[e^{a}_{m}, \psi^{a}_{\alpha}, b^{a}, M, A, \chi_{\alpha}, F]    
+ \mathcal{L}_{M}[e^{a}_{m}, \psi^{a}_{\alpha}, b^{a}, M, \Psi] \,,
\label{co}
\eeq
where we introduced the gravitino $\psi^a_{\alpha}$ and two auxiliary fields:
a vector $b^a$ and a scalar $M$. Eq.~(\ref{co}) is the supersymmetric version of eq.~(\ref{bd}).
The crucial difference between the two Lagrangians is that in the supersymmetric one
$\mathcal{L}_{M}$ and $\mathcal{L}_{\Phi}$ communicate 
also through the auxiliary fields $b^{a}$ and $M$. 
This has deep phenomenological consequences when the auxiliary fields are removed by means of their equations of motion.
To show this point, let us write the general solution of the equations of motion for $M$ and $b^{a}$ as follows 
\beqra
b^{a} &=& h_{1}(\dots, A,\chi_{\alpha}) \,, \nonumber\\
M &=& h_{2}(\dots, A,\chi_{\alpha}) \,,
\label{bm}
\eeqra
where $h_{1}$ and $h_{2}$ are two appropriate functions of the fields involved in the theory.
In eq.~(\ref{bm}) we underlined the crucial dependence of $h_{1}$ and $h_{2}$ from $A$ and $\chi_{\alpha}$. 
Now, replacing the solutions~(\ref{bm}) in the Lagrangian~(\ref{co}), the degrees of freedom of the
Planck multiplet explicitly appear in the matter Lagrangian. Since no Weyl rescaling of the vielbein
can remove the auxiliary fields from $\mathcal{L}_{M}$,  it follows that the Planck
multiplet couples intrinsically to matter. 
Therefore, there is no way to write the matter Lagrangian as 
$ \mathcal{L}_{M}[e^{a}_{m}, \psi^{a}_{\alpha}, \Psi]$ by means of a 
suitable vielbein redefinition of the form 
$e^{a}_{m} \to  e^{a}_{m}\, e^{l(A,\chi_{\alpha},F)}$, where $l$ is an appropriate function
of the components of $\Phi$.
In other words, a Jordan frame does not exist for such a theory.
The main consequence is that in the MSBD theory
the weak equivalence principle is not satisfied by construction and time variations
of masses and couplings are not under control.  
Explicit expressions for eqs.~(\ref{co}) and~(\ref{bm}) can be found in \cite{Riccardo}.

%%%%%%%%%%%%%%%%%%%%%%%%%%%%%%%%%%%%%%%%%%%%%%%%%%%%%%%%%%%%%%%%%%%%%%%%%%%%%%%%%%%%%%%%%%%%%%%%%
\section{Conclusions}
\label{com}
%%%%%%%%%%%%%%%%%%%%%%%%%%%%%%%%%%%%%%%%%%%%%%%%%%%%%%%%%%%%%%%%%%%%%%%%%%%%%%%%%%%%%%%%%%%%%%%%%

In this work we have studied the minimal supersymmetric extension 
of the BD theory (MSBD) defined by eq.~(\ref{SBD}).
The underlying motivation was the research of possible connections between a Scalar-Tensor interpretation
of Dark Energy and a supersymmetric description of Dark Matter.  
Eq.~(\ref{SBD}) is obtained replacing the Planck mass with a chiral superfield
in the supergravity Lagrangian~(\ref{sg}). We called this extra superfield the 
Planck superfield.
Although this approach looks very natural, the resulting phenomenology
is radically different from the one of the original BD theory.
In the MSBD theory the extra degrees of freedom of the Planck superfield
intrinsically couple to matter and a Jordan frame formulation can not be achieved
through a suitable vielbein redefinition.
As a consequence, this theory does not satisfy the weak equivalence principle by construction.
This conclusion could make the minimal supersymmetric extension of the BD idea phenomenologically inconsistent.
%\footnote{Here by ``inconsistent'' we mean that the weak equivalence principle is not satisfied
%by construction. For any other possible inconsistency or constraint that apply to ST theories, 
%see for instance \cite{Dam} and references therein.}
%
%In spite of this result, we find that if a consistent supersymmetric Scalar-Tensor theory were constructed, 
%it could provide a natural framework to achieve a Dark Matter-Dark Energy unification.
%For instance, in such a scenario Dark Matter and Dark Energy could be identified
%with different components of the Planck superfield. This issue is at present under analysis. 

%%%%%%%%%%%%%%%%%%%%%%%%%%%%%%%%%%%%%%%%%%%%%%%%%%%%%%%%%%%%%%%%%%%%%%%%%%%%%%%%%%%%%%%%%%%%%%%%%%
\section*{Acknowledgments} 
%%%%%%%%%%%%%%%%%%%%%%%%%%%%%%%%%%%%%%%%%%%%%%%%%%%%%%%%%%%%%%%%%%%%%%%%%%%%%%%%%%%%%%%%%%%%%%%%%%%
I sincerely thank Massimo Pietroni for many useful 
suggestions and discussions on Scalar-Tensor theories and their possible supersymmetric 
extensions. I would also like to thank Wilfried Buchmueller for
interesting discussions on the topic.
% and Massimo Pietroni and Gonzalo Palma 
%for having read and commented on a draft of the paper.
I finally acknowledges a Research Grant funded by the VIPAC Institute.

%%%%%%%%%%%%%%%%%%%%%%%%%%%%%%%%%%%%%%%%%%%%%%%%%%%%%%%%%%%%%%%%%%%%%%%%%%%%%%%%%%%%%%%%%%%%%%%%%

\end{document}